# Global Mapping of Gene/Protein Interactions in PubMed Abstracts: A Framework and an Experiment with P53 Interactions

Xin Li[1], Hsinchun Chen[1], Zan Huang[2], Hua Su[1], and Jesse D. Martinez[3]

[1]Artificial Intelligence Lab, Department of Management Information Systems, The University of Arizona, McClelland Hall, 1130 East Helen Street, Tucson, AZ 85721-0108, USA

[2]Department of Supply Chain and Information Systems, Smeal College of Business, The Pennsylvania State University, University Park, PA 16802, USA

[3]The Arizona Cancer Center, The University of Arizona, 1515 North Campbell Avenue, Tucson, AZ 85724, USA

Xin Li
1130 E Helen St, Room 430
Tucson, AZ 85721
xinli@email.arizona.edu
FAX: 520-621-2433

**Abstract**

Gene/protein interactions provide critical information for a thorough understanding of cellular processes. Recently, considerable interest and effort has been focused on the construction and analysis of genome-wide gene networks. The large body of biomedical literature is an important source of gene/protein interaction information. Recent advances in text mining tools have made it possible to automatically extract such documented interactions from free-text literature. In this paper, we propose a comprehensive framework for constructing and analyzing large-scale gene functional networks based on the gene/protein interactions extracted from biomedical literature repositories using text mining tools. Our proposed framework consists of analyses of the network topology, network topology-gene function relationship, and temporal network evolution to distill valuable information embedded in the gene functional interactions in literature. We demonstrate the application of the proposed framework using a testbed of P53-related PubMed abstracts, which shows that literature-based P53 networks exhibit small-world and scale-free properties. We also found that high degree genes in the literature-based networks have a high probability of appearing in the manually curated database and genes in the same pathway tend to form local clusters in our literature-based networks. Temporal analysis showed that genes interacting with many other genes tend to be involved in a large number of newly discovered interactions.

## 1  Introduction

Biological research has made it clear that cellular processes are controlled by interactions between genes, proteins, and other molecules. Detailed characterization of interactions between individual genes or proteins has been one of the focuses of traditional biological research. A new area known as network biology, which can be attributed to the recent advances in genomic technology, has emerged and many studies have tried to construct and analyze gene/protein interaction networks at a genome/proteome-wide scale to describe their global characteristics (1). (Gene/protein interactions in this paper include interactions between two genes, two proteins, or between a gene and a protein.)

Most studies in network biology rely on large-scale experimental data or manually collected knowledge to construct the networks. However, these studies are limited by the noise in experimental data and the intensive labor required for manual compilation of data. Previously, Barabasi and Oltvai (1) suggested using more advanced experimental tools for better biomedical interaction identification and quantification. Sharom et al. (2) proposed integrating different kinds of experimental datasets for better performance.

Biomedical literature, reliably and frequently documenting gene/protein interactions, can be an important data source for studying gene/protein interaction networks. Recent advances in text mining techniques make it possible to utilize high-coverage biomedical literature repositories, such as PubMed, to automatically extract gene/protein interactions and construct the corresponding networks. Such literature-based networks are valuable for characterization of the accumulated knowledge regarding gene/protein interactions. They also represent the collective human effort in knowledge exploration over a relatively long time span.

In this paper, we propose a framework for constructing and analyzing gene/protein interaction networks automatically extracted from biomedical literature. In this framework, we map the proteins to their encoding genes and study the interaction network at the gene level. We refer to this kind of abstract interaction, which contains both gene and protein interaction information, as gene functional interaction and these networks as gene functional networks (3). We demonstrate the application of our mapping framework using literature abstracts extracted from PubMed that are relevant to the gene P53 (a central player in cell cycle regulation and cancer development) as our testbed.

## 2 Background

Cellular regulatory pathways and networks that consist of gene functional interactions control many important biological processes in a cell (4). As an important topic in system biology in general, numerous efforts have been made to construct gene functional networks using different types of information sources (3). Understanding and analyzing such gene functional networks holds great potential to untangle the complexity of the underlying cellular processes (1, 2). Network visualization provides an intuitive presentation of gene interaction relations that allows researchers to easily understand the network structure of the relationships. It also enables the researchers to perform a wide range of information exploration tasks much more effectively and efficiently than a textual presentation (5). However, network visualization is usually more effective with relatively small size networks. This is due to the limitations of visualization algorithms and screen size and more importantly the human cognitive capabilities. For networks with hundreds of nodes, it is typically difficult to capture the structural properties visually. To understand the global structure of large-scale gene functional networks and other biological networks, network topological analysis methods have been applied in biomedical research.

Network topological analysis employs various statistical measures to characterize the topology of a large-scale complex network. These measures describe the important quantitative features such as the distance between nodes (average path length), tendency for the nodes to form clusters (clustering coefficient), and node degree distribution. Three important random graph models, the Erdos-Renyi model (6), the small-world model (7), and the scale-free model (8), have been the major analytical tools for understanding the governing principles of network topology. Recent empirical literature shows that the models could describe topological characteristics across a wide range of natural, social science, and technical networks (9).

Network topological analysis has been applied in many studies of various types of biological networks. We briefly review related studies on network biology and propose a taxonomy that characterizes biological network analysis in three dimensions: network types, data sources, and research focuses.

Based on the different levels of integration of cellular processes, the biological networks can be classified into four types: gene interaction networks, representing genome (or transcriptome)-wide interactions (10-12); protein interaction networks, representing proteome-wide interactions (13, 14); signal transduction networks, for interactions between genes, proteins, and other cellular signaling molecules (15, 16); and metabolic networks, for biochemical interactions between substrates and enzymes (17).

Biological networks can be constructed based on different types of data sources using a variety of analytical methods. High-throughout experimental data, such as microarray (10, 12), data from mass spectrometric analysis (18, 19), and two-hybrid screening (13, 20), is widely used in constructing gene or protein interaction networks. The existence of signaling and biomedical interactions can be determined using various analytical methods, including gene coexpression (21), transcriptome similarity (22), mutation screening (12), and so forth.

**Table 1.** A taxonomy for biological network analysis study

| Dimension | Type | Description | Examples |
|---|---|---|---|
| Network Types | Gene interaction networks | Networks represent the interactions at the gene level | S. cerevisiae (Tong et al., 2004 ; van Noort et al., 2004; Luscombe et al., 2004) Mammalian (Shaw, 2003) P53 (Hallinan 2004) |
| | Protein interaction networks | Networks represent the protein interaction relationship | S. cerevisiae (Jeong et al., 2001; Wuchty et al., 2003; Yook et al., 2004) |
| | Metabolic networks | Networks represent the relationship of the substrates in the same metabolic pathway | E. coli (Fell and Wagner, 2000; Wagner & Fell, 2001) 43 organisms (Jeong et al., 2000; Ravasz et al., 2002) 65 organisms (Ma & Zeng, 2003) |
| | Signal transduction networks | Networks represent the for interactions between genes, proteins, and other cellular signaling molecules. | S. cerevisiae (Luscombe et al., 2004) E. coli (Shen-Orr et al., 2002) Cancer protein (Jonsson, 2006) |
| Data Sources | Experimental data | Relations or correlations derived from the experimental data | Two-hybrid (Jeong et al., 2001; Yook et al., 2004) Microarray (Shaw, 2003; Tong et al., 2004; Luscombe et al., 2004; Carter et al., 2004; Noort et al., 2004) |
| | Manually compiled ontology or knowledge base | Interactions curated by experts based on prior knowledge | GO (Tari 2005) Manually compiled (Shen-Orr et al., 2002; Hallinan, 2004; Fell and Wagner, 2000; Wagner & Fell, 2001) Knowledge base (Ma and Zeng 2003; Wuchty et al., 2003; Yook et al., 2004) |
| | Literature-based data | Relations parsed using NLP tools or co-occurrence tools | Genes parsed from abstracts searched from PubMed by some keywords (Chen and Sharp 2004) |
| Research Focus | Topological characteristics | Topological measures and models | Small-world (Fell and Wagner 2000; Tari 2005) Scale-free (Jeong et al., 2001; Yook et al., 2004; Wagner and Fell, 2001; Shaw, 2003; Tari 2005) Hierarchical structure ( Ravasz et al., 2002;) Giant strong component (Ma and Zeng, 2003) |
| | Local structures | Special local structures and clusters | Network motif (Luscombe et al., 2004; Wuchty et al., 2003; Shen-Orr et al., 2002) High tendency to cluster (Tong et al., 2004; Carter et al., 2004) |
| | Topology-function relationship | Correlation between topological characteristics and biological functions | High degree node -> essential (Jeong et al., 2000) Correlation between network structure and protein function and location (Yook et al., 2004) Small world -> central metabolites (Ma and Zeng, 2003) Small world -> gene evolution (Noort et al., 2004) Clusters -> gene pathway (Hallinan, 2004) |

Manually curated ontologies or knowledge bases are created by domain experts based on previous research and literature. In some research, the biological interactions documented in knowledge bases, such as molecule reactions, are directly used in the construction of the biological network (17, 23, 24). Other research uses relations defined by an ontology, e.g., genes in the same GO functional group are considered related to each other (25).

Biomedical literature is another resource from which biological interactions can be extracted using statistical or Natural Language Processing (NLP) methods. The extracted interactions often take the form of binary relations between entities such as genes, proteins, or substrates. Most current studies try to map the entity co-occurrence relations in literature to biological relations (26). Chen and Sharp developed a system which incorporates NLP tools to parse syntactic gene relations from the searched PubMed abstracts using keywords. They reported the gene degree distribution of some parsed relation network examples (27).

Biological network analysis research focuses on three areas: network topological characteristics, local structures, and topology-function relationships.

- Research on network topological characteristics describes the global structure of the biological network by topological measures and models. Small-world and scale-free models have been widely used to describe the structure of gene interaction networks (25, 28), protein interaction networks (13, 20), signal transduction networks (29), and metabolic networks (17, 30, 31). A hierarchical structure model has also been proposed (32) to describe the structure of metabolic networks and other complex networks.

- Research on network local structures focuses on the common characteristics among a subset of closely related genes or proteins. Many studies have discovered that network motifs, i.e., recurrent interconnection patterns, exist in gene interaction networks (10, 23) and protein interaction networks (14).

- Research on topology-function relationships investigates the correlation between certain biological functions and network topological characteristics. Jeong et al. (30) found that high degree genes in a gene interaction network are more essential in cellular processes. It has also been found that genes in the same pathway (33), proteins in the same function group (20), or the same cellular localization (20) have a higher chance of interacting and forming clusters.

Table 1 summarizes the above dimensions with some examples.

Previous analyses have identified several important topological characteristics of different types of biological networks based on experimental data and manually curated data. Experimental data can provide a complete coverage of the genome (often tens of thousands of genes), but it contains a significant amount of noise and is limited to particular experimental conditions. Manually curated data is noise-free, but it requires intensive labor by domain experts. With the rapid development of biomedical research it has become even more difficult to collect biological interactions manually.

Using modern text mining techniques to automatically extract gene/protein relations from a large body of biomedical literature could be another way to construct gene functional networks. The biological literature documents the most important discoveries and provides an abundant resource of gene functional relation information, which has been validated by the experiments conducted by the authors and checked by the reviewers. Such biomedical literature is a large-scale resource for high-quality gene interaction information.

As an example of biomedical literature repositories, PubMed had collected about 16 million articles by the end of 2005 and hundreds of newly published articles are added to the collection every day. The scale of the biomedical literature necessitates the application of text mining techniques to automatic information extraction. Currently several tools have been developed to automatically extract gene/protein entities, gene/protein functions (34), and gene/protein interactions (35) from literature, but few of them use automatically extracted biological information to study the topological characteristics of gene functional networks. On the other hand, network information automatically extracted from text has been investigated in a wide variety of other domains, such as co-authorship networks (36, 37), citation networks (38, 39), and word adjacency networks (40).

There are two general types of gene functional relations extracted from literature: co-occurrence relations and parsed relations. Co-occurrence relations, which represent the appearance of two entities in the same context, are one way to represent gene/protein interactions (41, 42). Although not every co-occurrence relation reflects an actual interaction between the two genes, statistically significant co-occurrence relations based on a large corpus of literature may correspond to underlying gene interactions. Parsing relations using Natural Language Processing (NLP) technology is another approach to gene/protein interaction extraction. McDonald et al. (43) classified the NLP approaches that are used in biological relation parsing into three categories: syntactic parsing (27); semantic parsing (44); and balanced approach, which use both sentences' syntactic information and entities' semantic information (34, 43, 45). These NLP approaches can achieve a high parsing accuracy in gene interaction extraction. For instance, the Arizona Relation Parser (43) achieved a precision of over 90% and a recall of over 60%. The advances in text mining tools make it possible to process large-scale biomedical literature and extract gene/protein interactions efficiently with acceptable accuracy.

One should note that due to inherent difficulty of text mining, it is difficult to achieve 100% accuracy in gene/protein relation extraction. It is also possible that the extracted relations do not represent the actual underlying gene functional interactions, wince experimental studies under different conditions may have resulted in conflicting relations. Previously well-documented relations may be proven to be incorrect by later studies. Thus literature-based networks can only play a supplementary role to biological experiments in identifying possible gene functional relations. Even with these drawbacks, literature-based gene functional networks provide a compact summary of previous gene functional relation discoveries and greatly alleviate the information overload problem faced by every biomedical researcher. Furthermore, we believe structural analysis of the literature networks can provide valuable insights into the underlying gene/protein regulation process as well as the biomedical knowledge creation and exploration process.

Most previous research on network biology was based on experimental data and manually curated functional relations. In this paper we propose a comprehensive framework for constructing and analyzing gene functional networks using co-occurrence relations and parsed relations obtained from automatic processing of a large-scale corpus of biomedical literature.

## 3   A literature-based gene functional network analysis framework

As shown in Figure 1, our framework consists of four major steps: document collection, gene/protein interaction extraction and aggregation, network construction/visualization, and network analysis.

**Figure 1.** Literature-based gene network analysis framework

## 3.1  Document collection

This step involves collecting of publications from major biomedical literature repositories. PubMed is the data source used in this research. Selected keywords are used to filter the literature abstracts that focus on a defined research area. Although both abstracts and full text articles can be used in our framework for gene/protein interaction extraction, we chose the abstracts in this research instead of full text due to several considerations: 1) Access to full text literature is often limited by copyright restriction (46), while most public databases provide access to literature abstracts. 2) Storing and processing large corpus of full text biomedical literature requires a large amount of computational capability (47). 3) Abstract usually has higher information density (48) and result in better quality relations extracted by text mining techniques. In general, the abstracts of biomedical literature often contain the most important findings and gene/protein interactions of the research article, while the full text articles may contain too much detail for the experiments and previous research findings. Considering the large number of abstracts we analyze, the resulting gene functional network may still have reasonably good coverage of the gene functional relations. We will leave the formal analysis of the difference between network construction using abstracts and full texts for future research.

## 3.2  Gene/protein interaction extraction and aggregation

In this component, we leverage the text mining techniques we previously developed to extract the gene/protein interactions from the collected PubMed abstracts. The process consists of two major steps: parsing gene/protein relations by the Arizona Relation Parser (43) and aggregating the parsed relations by the BioAggregate tagger (45).

The Arizona Relation Parser (ARP) is a generic relation parser which uses both syntactic and semantic rules to extract relational triples out of biomedical abstracts. A relational triple consists of two labeled entities and a labeled connector. If the sentence contains negation words, it will be noted in the negation tag. The ARP splits the sentences and tags the entities. It then uses heuristics and grammar knowledge to parse the relations. Semantic knowledge based on the UMLS and Gene Ontology is used to correct the tagging errors and construct the relational triples. Table 2 gives some examples of ARP output.

**Table 2.** Arizona Relation Parser output

| Original sentence | Resulting relation | | | |
| --- | --- | --- | --- | --- |
| | Entity 1 | Neg. | Connector | Entity 2 |
| The transcription of mdm2 gene is activated by P53 | P53 | False | activated | transcription of mdm2 gene |
| induction of P53 transcriptional activity leads to increases in mdm2 RNA | induction of P53 transcriptional activity | False | leads to increases | mdm2 RNA |
| high TP53 mRNA expression […] also had elevated levels of MDM2 mRNA | high TP53 mRNA expression | False | also had elevated | levels of MDM2 mRNA |

The BioAggregate tagger is used to re-process the relational triples extracted by the ARP into aggregated relations. In this process, the entities that represent the same gene with different names are aggregated into the same identifier. If an entity represents a gene product, it will be aggregated into the identifier of the gene. The connectors are aggregated into one of the four types of gene functional relations: activation, inhibition, directional, and unspecified association.

Existing gene and protein nomenclature and ontology sources, including RefSeq, Locuslink, HUGO, SGD, and Gene Ontology, are employed to create an Aggregatable Substance Lexicon and a Feature Lexicon. Then a decompositional tagger processes the triples using the lexicons and transforms them into aggregated relations. After this process, the interrelationships between genes and gene products are transformed into gene functional relations in a unified form. For example, the three relation triples in Table 2 are aggregated into "*P53 activates MDM2.*"

The precision and recall of the two tools have been tested in previous research. Compared against a benchmark of manually extracted relations, the ARP had a precision of 90.8% and a recall of 61.0%. In identifying genes and interactions in ARP relations, the BioAggregate Tagger had a precision of 81.7% and a recall of 51.4% (43, 45). Although the average recall is not very high, the overall recall of the whole network can be improved when the large volume of biomedical literature is taken into account. The same relation may appear multiple times in different papers and the probability that a relation is missing during the extraction process decreases significantly with multiple occurrences in the entire literature.

### 3.3   Network construction/visualization

After the relations are extracted, they are used to construct two types of literature-based gene functional networks (the gene products are mapped to the encoding genes): a *Parsed Network* consisting of the parsed gene functional relations and co-occurrence networks consisting of gene/protein co-occurrence relations.

To construct the *Parsed Network*, we include only aggregated relations with both entities recognized as genes or gene products. Each aggregated gene functional relation has a time tag which indicates the first time it was documented in biomedical literature. We exclude the relations with a true negation tag because they do not directly reflect the connection between genes in reality. For our study we also disregard the direction and connector type of the parsed interactions.

For the co-occurrence networks, two aggregated entities (as identified in the aggregation process) are considered connected if they appear in the same abstract. Using the co-occurrence relations extracted from a large body of abstracts, we can generate a co-occurrence network in which the existence of a link indicates that two genes or gene products co-occurred in at least one abstract. As co-occurrence relations appeared multiple times are more meaningful, we create a *Reduced Co-occurrence Network* by only including the high-frequency co-occurrence relationships and reduce the network to the same scale as the *Parsed Network*. In our research we focus on analyzing the *Reduced Co-occurrence Network,* while the results of the original co-occurrence network are also reported. The latter is referred to as the *Co-occurrence Network* hereinafter.

In this research, we use the manually curated TransPath database (49) as a benchmark. The TransPath database contains the reactions of genes and gene products. We map all the gene products to the corresponding genes and study the TransPath regulatory pathways at the gene level. The network contains the gene functional relations validated by domain experts, which we called the *Curated Network*. In our study, we treated the relations in the *Curated Network* as the correct underlying relations of the P53 pathway in order to assess the information quality of the different types of literature networks included in our framework. In practical applications, it is also possible that the relations extracted from large corpus of biomedical literature can supplement the existing manually curated databases of genetic relations or even identify and correct the errors in these databases.

The three networks can be visualized and analyzed using network visualization tools, such as GeneScene visualizer (50) or Cytoscape (5). Such tools can be used by biologists to inspect the details of the network. In this research, we focus more on analyzing the global structure of the networks.

## 3.4   Network analysis

This part of our work addresses the characteristics of the gene functional networks constructed from the literature. It consists of three components: topological analysis, topology-function relationship analysis, and temporal analysis. As the gene functional relations extracted from literature represent experimental results under various conditions and treatments in different species and tissue or cell types, the resulting network is a comprehensive picture overlaid with information from various sources rather than a precise snapshot of any particular cellular settings. This kind of integrated networks may enable researchers to identify missing links or unknown interactions in a given biological system and can potentially facilitate the processes of hypothesis development and new knowledge discovery.

### 3.4.1   Topological analysis

For network topological analysis, we apply several topological measures to the inference of the underlying mechanisms governing the gene functional network. The main topological measures we adopt are described as follows (1, 9):

(1) Average path length $l$: The average value of the shortest path lengths between any pair of nodes in the network.

(2) Network diameter $D$:  The maximum value of the shortest path lengths between any pair of nodes in the network.

(3) Clustering coefficient $C$: The network's clustering coefficient $C$ is the average of each node's clustering coefficient $C'$. A node's clustering coefficient is the ratio of the number of edges between the node's neighbors to the number of possible edges between those neighbors.

$$C' = \frac{\text{number of edges between the neighbors}}{\text{possible number of edges between the neighbors}}$$

(4) Average degree $<k>$: The average number of links that a node has to other nodes.

(5) Degree distribution $P(k)$ : Degree distribution gives the probability that a selected node has exactly $k$ links.

$$P(k) = \frac{N(k)}{N}$$

$N(k)$ is the number of nodes with k links; $N$ is the total number of nodes.

A network may contain several components. A component is an isolated sub-network in a disconnected network. A node in one component can reach any node in the same component but cannot reach a node outside the component. We also measure the number of components ($N_C$) of the network.

### 3.4.2 Topology-function relationship analysis

In this part, we study the relationship between some topological characteristics and their biological implications. We investigate the relationship between the gene degree in the literature-based networks and the validation of the gene's relation to a domain. We study the relationship between gene clusters and gene pathways in the literature-based networks.

Previous research on experimental data shows that high-degree nodes in a gene functional network are more essential (13, 21). We consider genes and gene functional relations in the manually curated benchmark *Curated Network* well-studied and validated. For each group of genes with the same degree in each literature-based network, the percentage of genes that are in the overlap part of *Curated Network* and the literature-based network is calculated. We hypothesize that high-degree genes have higher probability to be existed in the validated gene group. In other words, there is a positive correlation between gene degree and the validation of the gene's relation to a domain.

As genes in the same pathway tend to have more interactions with each other, they may form a closely related local cluster. To study the relationship between network clusters and gene pathways, we use an clustering algorithm proposed by Newman and Girvan (N-G algorithm) (51) to find clusters in the network, because of its reported good clustering performance in utilizing network topological information in network node clustering (52). In this algorithm, a network measure, link betweenness, is used as the distance between genes. The links are removed from the network one by one according to their link betweennesses in descending order. Each time a link is removed, the link betweenness measure of the network has to be recalculated. After removing some links, the network's cluster modularity $Q = Tr(E) - \|E^2\|$ is calculated ($E = \{e_{ij}| e_{ij}$ is the fraction of the links that link nodes in cluster $i$ with nodes in cluster $j\}$. *Tr(E)* is the trace of matrix $E$. $\|E^2\|$ is the sum of the elements of matrix $E^2$.) (51). Among all the cluster partition solutions, the one with the highest cluster modularity is selected, so that the number of its intra-cluster links is maximized and the number of its inter-cluster links is minimized. The clustering result is evaluated by the biologists at the Arizona Cancer Center in terms of correspondence between the clusters and gene pathways.

*3.4.3   Temporal analysis*

As the gene functional relations in the *Parsed Network* are stamped with the time of their first appearance in the literature, the temporal dimension of the network documents the discovery history of gene functional relations. We study the temporal characteristics of the *Parsed Network* by year. For each year *T*, a sub-network of the *Parsed Network* is studied, which includes all gene functional relations reported in abstracts published before the end of year *T*. The *Parsed Network*'s temporal evolution may provide valuable insight on the knowledge exploration process in biomedical domains.

Although gene/protein co-occurrence relations also have a timestamp indicating the time of the gene pair's first co-occurrence, such a timestamp is not a reflection of the discovery of the gene interaction. Thus we do not study the temporal co-occurrence network.

## 4  The P53 testbed

The well-studied P53 tumor suppressor gene plays a central role in the regulation of apoptosis and cell cycle arrest in cancer development. Because of the wide interest and rich literature on P53, we created a P53-related testbed to demonstrate the application of our framework. By including PubMed abstracts that contain various names of P53 and other genes in the P53 pathways, we identified a testbed of 87,903 abstracts (1975 – 2003).  From these abstracts, we extracted 51,033 distinct entities and 44,864 relational triples using the ARP. The parsed relations were aggregated into 4,233 genes and 33,968 relations, which are used to construct a *Co-occurrence Network*. Among these relations, the 6,875 non-negation interactions with genes on both sides, which contain 2,045 genes, are used to construct a *Parsed Network*. After pruning the *Co-occurrence Network* to the same scale as the *Parsed Network*, we obtained a *Reduced Co-occurrence Network* with 2,017 genes and 10,104 relations which contains the co-occurrence relations that appeared in two or more abstracts.

We constructed the *Curated Network* based on the TransPath database (version 5.1), which contains 17,054 reactions in various regulatory pathways. In these reactions, we identified the reactions whose reactants and products are either human genes or their products and mapped the reactions to the relations at gene level. The abstract network, which contains 657 relations between 454 human genes, is defined as the *Curated Network*. This network serves as a benchmark of well-recognized, true gene functional relations, curated by human experts.

## 5 Empirical results and discussion

### 5.1 Topological analysis

The topological measures of the four gene functional networks, *Curated Network*, *Parsed Network*, *Co-occurrence Network*, and *Reduced Co-occurrence Network,* are shown in Table 3. Degree distributions of these networks are shown in Figure 2.

All four networks are composed of several components. They all have a giant component which has most of the genes in the network. For example, the giant component of the *Parsed Network* contains 96% (1,967/2,045) of the nodes and 99% (6,050/6,092) of the links. Components are separated sub-networks in which all gene pairs are connected by at least one path. Thus, one gene may affect other genes in the same component through the gene interactions. The existence of the giant components, which is also found in other biological networks (24), indicates a high degree of interdependency between the genes involved in cellular processes.

**Table 3.** Topological analysis of gene/protein interaction networks

| Networks | *Curated Network* | *Parsed Network* | *Co-occurrence Network* | *Reduced Co-occurrence Network* |
|---|---|---|---|---|

| *Nodes* | 454 | 2045 | 4233 | 2017 |
|---|---|---|---|---|
| *Links* | 657 | 6092 | 33,968 | 10,104 |
| $<k>$ | 2.894 | 5.958 | 16.050 | 10.019 |
| $l$ | 4.441 | 3.318 | 2.884 | 2.891 |
| $l_{rand}$ | 5.757 | 4.271 | 3.009 | 3.302 |
| $C$ | 0.0459 | 0.3149 | 0.6254 | 0.6769 |
| $C_{rand}$ | 0.0064 | 0.0029 | 0.0038 | 0.0049 |
| $D$ | 11 | 8 | 8 | 8 |
| $N_c$ | 21 | 37 | 51 | 30 |
| $Node_c$ | 401 | 1967 | 4125 | 1956 |
| $Link_c$ | 624 | 6050 | 33903 | 10071 |

*Nodes*: Node number; *Links*: Link number; $<k>$: Average degree; $l$: Average path length; $C$: Clustering coefficient; $D$: Network diameter; $l_{rand}$: Average path length for the same size random network; $C_{rand}$: Clustering Coefficient for the same size random network; $N_c$: Number of components; $Node_c$: Number of nodes in the largest component; $Link_c$: Number of links in the largest component and degree distribution

Table 3 shows that all four networks have a large clustering coefficient and a small average path length compared to random networks of the same size. For example, the *Parsed Network* has a much larger clustering coefficient (0.3149) and a smaller average path length (3.318) than those of a same-size random network (0.0029 and 4.271, respectively). These properties reflect the small-world characteristics of the networks. A small average path length indicates that one gene's effect can be quickly propagated to other genes in the biological process. A large clustering coefficient indicates that the genes interacting with one gene tend to interact among themselves as well. In other words, there is a probability of the existence of local clusters.

From Table 3 we can observe that there is a major difference in the size of the *Reduced Co-occurrence Network* and the *Co-occurrence Network*, but their average path length and clustering coefficient are quite similar. In the *Co-occurrence Network* genes appearing in the same abstract form a fully connected cluster, and the network is made up of those local clusters. Thus the *Co-occurrence Network* has a high clustering coefficient and a small average path length. The Reduced Co-occurrence Network is formed by removing weak co-occurrence relations that only occurred in one abstract, which might not represent an actual gene functional relationship. The similarity in the topological measures of the two networks indicates that removing the rarely appearing co-occurrence relations from the network does not substantially change the network topology.

Although the four networks have similar average path lengths, there is a large difference in their clustering coefficients. The clustering coefficients of the *Reduced Co-occurrence Network* (0.6769) and the *Co-occurrence Network* (0.6254) are about twice as large as that of the *Parsed Network* (0.3149) and 20 times larger than that of the *Curated Network* (0.0459). The substantial difference in clustering coefficients reflects the nature of the three different networks in local cluster (highly connected sub-graph) formation. The *Curated Network* has the relations carefully collected by domain experts. The most notable and important relations are documented, which we refer to as gene pathways. The small clustering coefficient indicates that the pathways in the *Curated Network* do not have significant local clusters. Comparing the *Reduced Co-occurrence Network* and the *Parsed Network*, we can see that although the two networks have similar numbers of genes, the *Reduced Co-occurrence Network* has a much larger clustering coefficient. This indicates that the *Reduced Co-occurrence Network* captured many more relations and has more significant local clusters..

In addition, based on the average degree measure, the density levels of the networks are different. Biologists from the Arizona Cancer Center reviewed some examples of the different networks consisting of the same group of genes. They found that the *Curated Network* only contains the most important interactions. The *Parsed Network* may provide a more complete gene functional network than the *Curated Network*. The two *Co-occurrence Networks* are close to fully-connected networks. The *Parsed Network* contains less noise than the *Reduced Co-occurrence Network*, which may potentially hide important interactions.

**Figure 2.** Degree distribution of the gene functional networks

Figure 2 shows that the degree distributions of the four networks are close to a straight line, indicating that they follow a power-law distribution. A power-law distribution means that the number of nodes with a certain degree in the network decreases quickly when the degree increases. The degree distributions of the *Parsed Network*, the *Reduced Co-occurrence Network*, and the *Co-occurrence Network* also show a heavy tail, corresponding to the group of genes with a very large degree. While the power-law degree distribution (revealing a scale-free property of the network) with a heavy tail was also reported in other types of large-scale networks, it seems to have a biological implication in our context. In our P53 testbed, there are only a few high-degree genes in the network, but they affect many other genes. These high-degree genes may be the central players in the P53 pathway and related cellular processes. For instance, TP53, JUN, FOS, MYC, TNF, IL6, IL1B, and MAPK8 are among the top ten genes with the largest degrees in both the *Parsed Network* and the *Reduced Co-occurrence Network*. Being transcriptional factors, cytokines, or kinases, these genes are actively involved in regulation of cell proliferation, differentiation, inflammation, apoptosis, or tumor development (53-55). Many of them respond to various cell stimuli and act as an integration point for multiple biochemical signals, which is consistently reflected by their high degree in the networks.

The scale-free characteristics of the networks in our research may have two causes. 1) The actual gene interactions in biological processes follow the power-law distribution. 2) As our data is on the discovered genes and relations in the literature, the scale-free characteristics might be a result of the collective knowledge creation and accumulation process of human beings—researchers tend to conduct research related to the known important genes. We will discuss this again in the temporal analysis section.

## 5.2 Topology-function relationship analysis

**Figure 3.** High-degree genes have higher probability to be documented in a human-curated knowledge base

### 5.2.1 Essential genes

As discussed before, genes with large degrees in a literature-based network of a particular domain (P53 related in our case) were documented to interact with many other genes and are likely to be central players of that domain. To further explore this relationship, we used the *Curated Network* as a benchmark to study the *Parsed Network* and two *Co-occurrence Networks* to see whether high degree genes are more likely to be documented in a human-curated knowledge base and validated by experts to be relevant to the P53 pathways. The result is shown in Figure 3 and Table 4. In Figure 3, we divided the genes into 11 groups according to their degrees. The percentages of validated genes (genes appeared the *Curated Network*) for the three literature-based networks are reported. The results show that the *Parsed Network* and the *Reduced Co-occurrence Network* has higher validated gene percentages than the *Co-occurrence Network* in most of the groups. From Figure 3, we can observe that high-degree genes in the literature-based networks are more likely to be validated genes. For example, for the genes with a degree between 1 and 5 in the *Parsed Network*, 1.67% are validated genes (i.e., genes appeared in the *Curated Network*), while for the genes with a degree larger than 50, 45% are validated genes. For these degree-based gene groups we observe a positive correlation between the group index (0 representing degree 1-5, 1 to represent degree 6-10, etc.) and the percentage of validated genes. This correlation (thereafter referred to as *degree-validation correlation*) reveals the quality of the literature network in terms of identifying essential genes using gene degrees. We calculated the Pearson correlation coefficient (13) between group index and percentage of validated genes for the three literature networks. As shown in Table 4, the degree-validation correlation coefficient of the *Parsed Network* (0.866) is the highest among the three networks, followed by the *Co-occurrence Network* (0.751). The degree-validation correlation coefficient of the *Reduced Co-occurrence Network* (0.590) is much smaller than the other two networks. These results show that a gene's degree in the *Parsed Network* is most informative regarding whether this gene is central to the P53 pathway comparing with the other two networks.

**Table 4.** High degree genes are essential

| Gene degree | Group index | Validated genes percentages | | |
| --- | --- | --- | --- | --- |
| | | *Parsed Network* | *Co-occurrence Network* | *Reduced Co-occurrence Network* |
| 1~5 | 0 | 1.67% | 0.68% | 1.71% |
| 6~10 | 1 | 4.50% | 1.29% | 5.58% |
| 11~15 | 2 | 8.05% | 4.33% | 5.74% |
| 16~20 | 3 | 5.26% | 5.85% | 10.67% |
| 21~25 | 4 | 14.29% | 8.55% | 11.11% |
| 26~30 | 5 | 11.76% | 9.64% | 11.11% |
| 31~35 | 6 | 11.11% | 5.08% | 30.77% |
| 36~40 | 7 | 10.00% | 5.71% | 10.00% |
| 41~45 | 8 | 33.33% | 7.69% | 0.00% |
| 46~50 | 9 | 28.57% | 9.52% | 16.67% |
| >50 | 10 | 44.83% | 24.37% | 42.25% |
| Correlation coefficient | | 0.866 | 0.751 | 0.590 |

### 5.2.2  Gene clusters

To study the relationship between cluster structures and gene pathways, we used the N-G algorithm (51) to identify gene clusters. As we mentioned in section 3.4.2, the link betweenness measure has to be recalculated each time we remove a link from the network. To find an appropriate cluster partition, this measure has to be calculated multiple times. Because of the intensive computational requirements of the clustering algorithm, we restricted the number of genes (relations) in the clustering process. For the *Parsed Network*, we only kept the interactions that appeared in more than one paper. We also further reduced the *Co-occurrence Network* to about the same size as the reduced *Parsed Network,* based on occurrence count. The reduced *Parsed Network* contains 618 genes and 2,704 relations. The further reduced *Co-occurrence Network* contains 848 genes and 2,984 relations, covering the co-occurrence relations appearing in three or more abstracts. The clustering algorithm identified 31 clusters from the reduced *Parsed Network* and 40 clusters from the further reduced *Co-occurrence Network*.

**Table 5.**  Gene clusters and gene pathways

| Dominant pathways of the cluster | *Parsed Network* Cluster | | | *Co-occurrence Network* Cluster | | |
|---|---|---|---|---|---|---|
| | Cluster Size | Related Genes | Related Genes % | Cluster Size | Related Genes | Related Genes % |
| P53/apoptosis pathway | 83 | 54 | 65.06 | 76 | 48 | 63.16 |
| Cytokine signaling & immune response | 74 | 59 | 79.73 | 107 | 76 | 71.03 |
| Cell cycle regulation & tumorigenesis | 53 | 39 | 73.58 | 53 | 39 | 73.58 |
| MAPK signaling pathway | 45 | 39 | 86.67 | 36 | 30 | 83.33 |
| Small GTPase mediated signaling | 24 | 22 | 91.67 | 10 | 8 | 80.00 |

Due to the difficulty in evaluating gene clusters, we conducted a small-scale evaluation study of gene clusters generated. We found that most clusters are related to one or two dominant pathways. Five generally consistent clusters appearing in each of the networks were reviewed by a Ph.D. researcher in biology. The clusters were compared with well-documented gene interactions and pathways in public resources, including KEGG, Entrez Gene, OMIM, and PubMed. Table 5 shows the accuracies of the clusters, defined as the percentage of related genes in the clusters (The related genes are relevant to the dominant pathway(s) according to the knowledge sources we used). The five clusters all have relative high accuracies, which imply that most genes that are clustered together are involved in the same pathway. The expert also found some "possibly related genes" in the clusters. Although there is no evidence of the direct interaction between these genes and the dominate pathways, they may interact with the genes in the pathway indirectly. These "possibly related genes" could provide hints to biologists for further studies that may lead to new discoveries.

## 5.3 Temporal analysis

The large-scale high-quality gene functional relations in the *Parsed Network* also allow us to perform temporal analysis on the evolution of the network of gene functional relations.

**Figure 4.** Literature Network size evolution

Figure 4 shows the evolution of the number of nodes and links of the *Parsed Network*. We observed consistent growth in the number of new interactions and genes, especially in recent years (after 1991). The decrease in the number of newly discovered gene functional relations in 2003 is because of incomplete data at the time of study. Except for that, there is no indication of the network's convergence to a fixed set of genes. It is possible that more genes involved in the P53 pathway will be identified.

**Figure 5.** Preferential attachment test

In Figure 5, the preferential attachment test (56) shows more details about the expansion of the *Parsed Network*. The preferential attachment tests for the years from 1976 to 2003 follow similar patterns. We only reported results for the most recent five years to make the graph easier to read. The straight line of cumulative preferential attachment $K(k)$ with positive slope in the log-log graph means that node $i$'s probability to get new links $P_i(k)$ is proportional to its degree $k$. Thus, the probability a gene will be found to interact with other genes is proportional to the number of known interactions involving it. This indicates that researchers tend to focus more on the well-studied genes and study the gene functional relations related to them. This analysis shows evidence that the nature of the collective research exploration process at least partially account for the observed power-law degree distribution of the literature-based networks reported earlier in Section 5.1.

## 6   Conclusions and future directions

In this paper, we described a comprehensive framework for constructing and analyzing large-scale literature-based gene functional networks. We focused on constructing literature-based networks using gene functional relations parsed from sentences and gene co-occurrence patterns in biomedical abstracts. Three types of useful analyses, including network topological analysis, topology-function relationship analysis, and temporal analysis, were conducted to analyze both types of literature-based networks. We demonstrated the application of this framework using a testbed of PubMed abstracts related to P53 pathways.

In the P53 dataset, the gene functional networks extracted from literature using the NLP approach (*Parsed Network*) and using the co-occurrence approach (*Reduced Co-occurrence Network*), and the gene functional network manually curated by domain experts (*Curated Network*) show similar topological characteristics. These networks all have small-world and scale-free properties. Comparison of the networks shows that the *Reduced Co-occurrence Network* contains more significant local clusters than the *Parsed Network*, while the *Parsed Network* contains less noise than the *Reduced Co-occurrence Network*.

We found that high-degree genes in the literature-based network are more likely to appear in the manually curated gene functional network. Genes in the literature-based networks' clusters are highly related to each other, many belonging to the same pathway or inter-linked pathways. The evolution of the *Parsed Network* shows preferential attachment characteristics, which is consistent with other large-scale networks.

In future research, we will extend the scope of our studies to other research topic areas such as cancer-related genes. We will combine gene functional networks from different sources to form a comprehensive network and study its topological features and evolution. We will also consider including the direction of the interactions in analysis.

## Acknowledgements

This work was supported by: NIH/NLM, 1 R33 LM07299-01, 2003-2005, "Genescene: a Toolkit for Gene Pathway Analysis." We would like to thank researchers at the Arizona Cancer Center for their helpful comments.

**Figure 1**

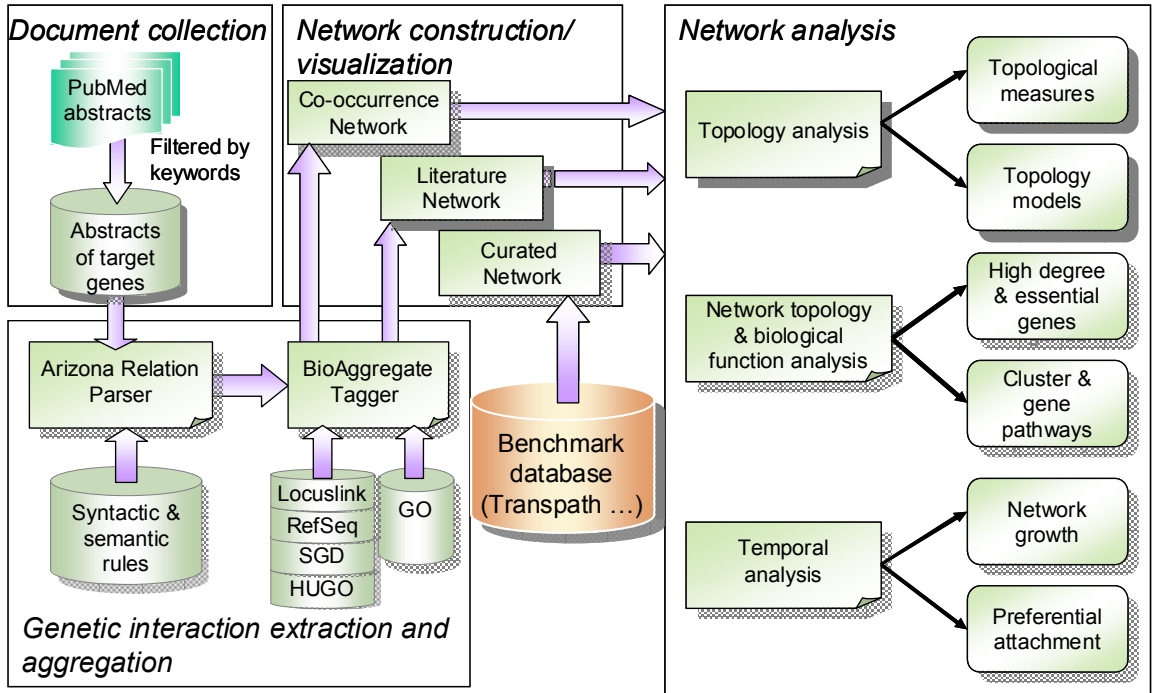

**Figure 1.** Literature-based gene network analysis framework

**Figure 2.** Degree distribution of the gene functional networks

**Figure 3**

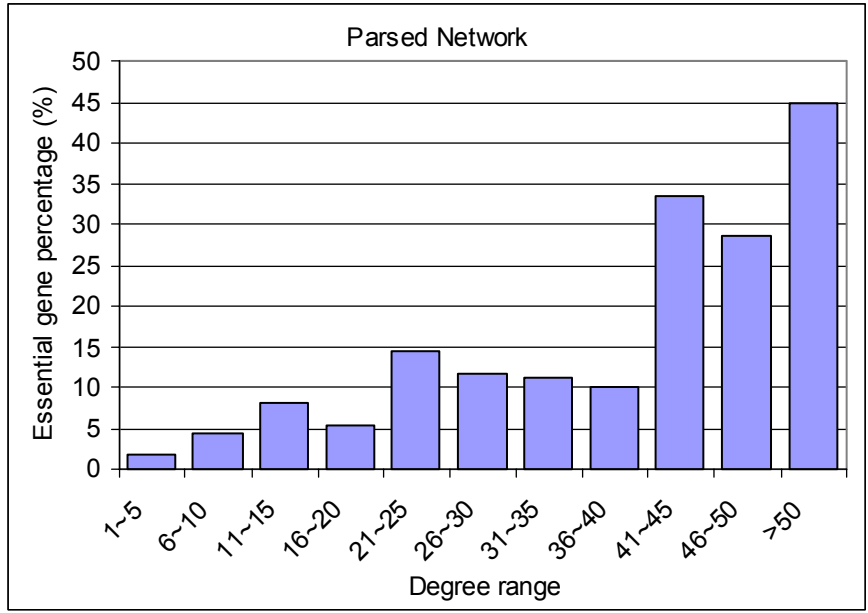

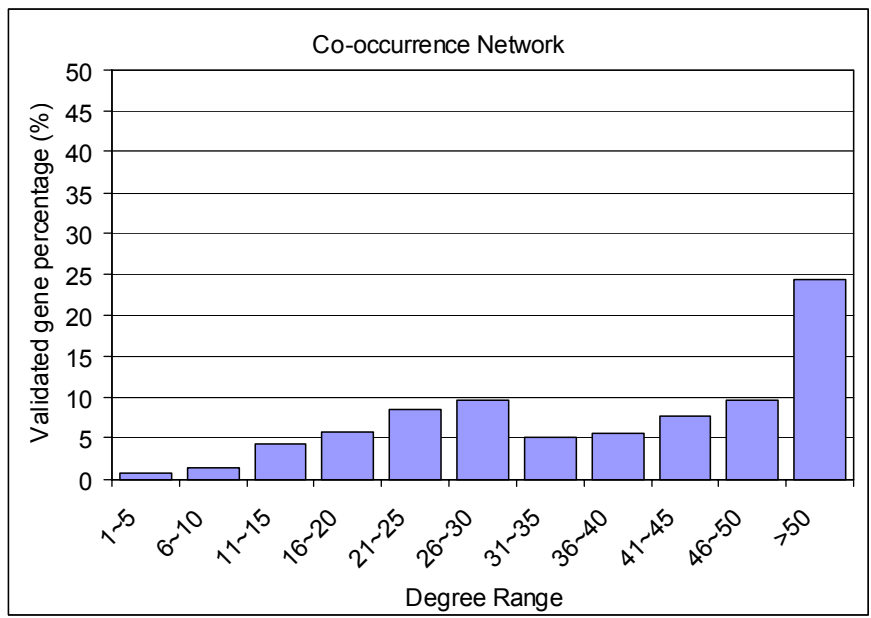

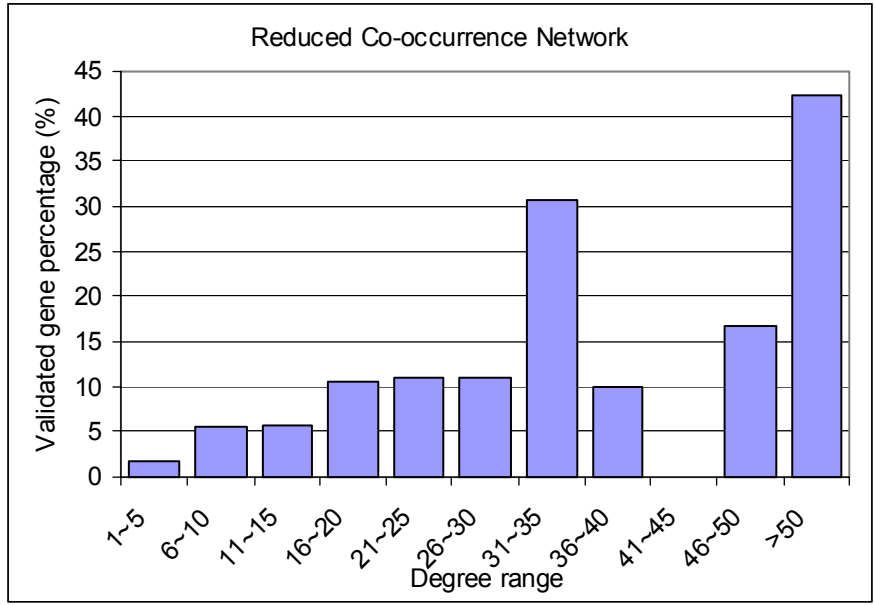

**Figure 3.** High-degree genes have higher probability to be documented in a human-curated knowledge base

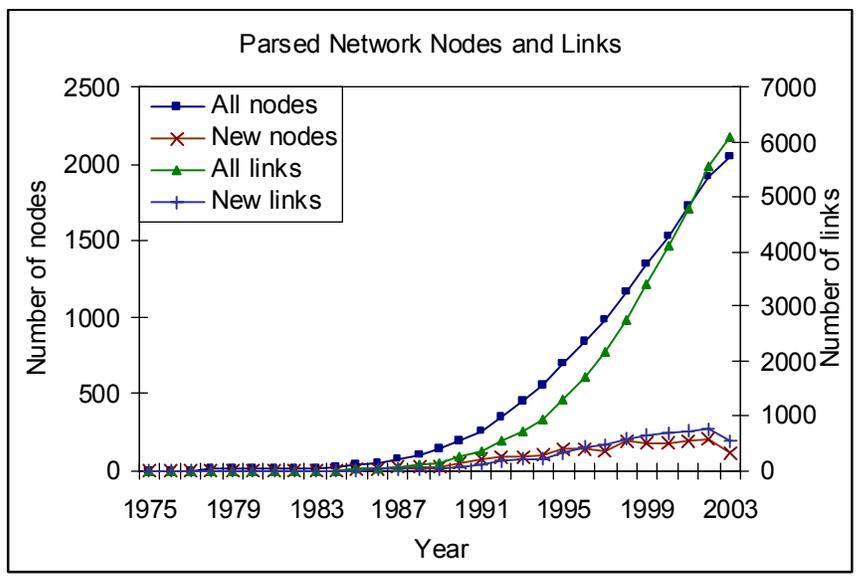

**Figure 4.** Literature Network size evolution

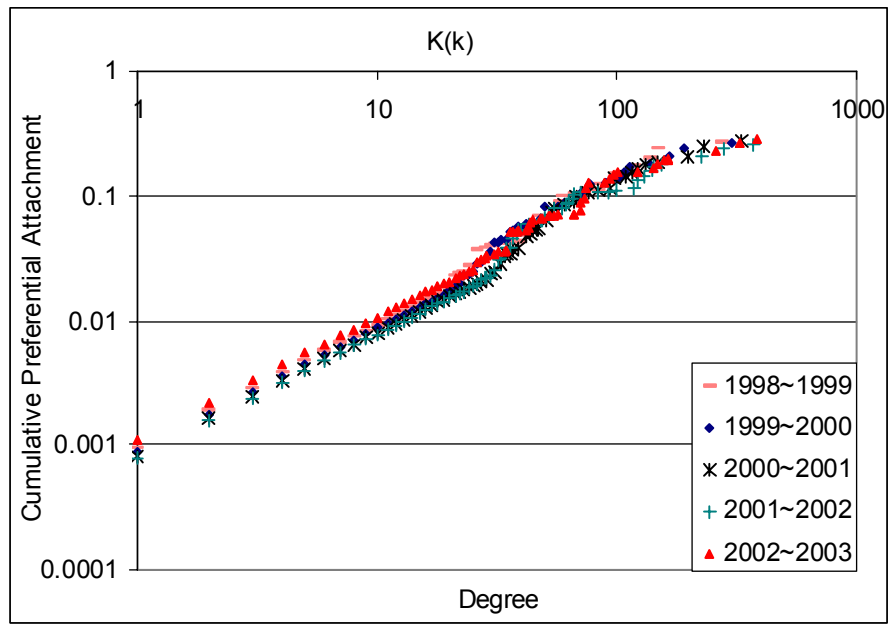

**Figure 5.** Preferential attachment test